\documentclass[twoside,twocolumn,9pt]{article}
\usepackage[utf8]{inputenc}
\usepackage{extsizes}
\usepackage[super,sort&compress,comma]{natbib} 
\usepackage[version=3]{mhchem}
\usepackage[left=1.5cm, right=1.5cm, top=1.785cm, bottom=2.0cm]{geometry}
\usepackage{balance}
\usepackage{times,mathptmx}
\usepackage{sectsty}
\usepackage{graphicx} 
\usepackage{lastpage}
\usepackage[format=plain,justification=justified,singlelinecheck=false,font={stretch=1.125,small,sf},labelfont=bf,labelsep=space]{caption}
\usepackage{float}
\usepackage{fancyhdr}
\usepackage{fnpos}
\usepackage[english]{babel}
\addto{\captionsenglish}{%
  
}
\usepackage{amsmath}
\usepackage{amssymb}
\DeclareMathOperator{\Tr}{Tr}
\usepackage{array}
\usepackage{droidsans}
\usepackage{charter}
\usepackage[T1]{fontenc}
\usepackage[usenames,dvipsnames]{xcolor}
\usepackage{setspace}
\usepackage[compact]{titlesec}
\usepackage{hyperref}
\usepackage{epstopdf}%This line makes .eps figures into .pdf - please comment out if not required.

\definecolor{cream}{RGB}{222,217,201}

\begin{document}

\pagestyle{fancy}
\thispagestyle{plain}
\fancypagestyle{plain}{

%%%HEADER%%%
% \fancyhead[C]{\includegraphics[height=18.5cm]{head_foot/header_bar}}
%  \fancyhead[L]{\hspace{0cm}\vspace{1.5cm}\includegraphics[height=30pt]{head_foot/journal_name}}
% \fancyhead[R]{\hspace{0cm}\vspace{1.7cm}\includegraphics[height=55pt]{head_foot/RSC_LOGO_CMYK}}
\renewcommand{\headrulewidth}{0pt}
}
%%%END OF HEADER%%%

%%%PAGE SETUP - Please do not change any commands within this section%%%
\makeFNbottom
\makeatletter
\renewcommand\LARGE{\@setfontsize\LARGE{15pt}{17}}
\renewcommand\Large{\@setfontsize\Large{12pt}{14}}
\renewcommand\large{\@setfontsize\large{10pt}{12}}
\renewcommand\footnotesize{\@setfontsize\footnotesize{7pt}{10}}
\makeatother

\renewcommand{\thefootnote}{\fnsymbol{footnote}}
\renewcommand\footnoterule{\vspace*{1pt}% 
\color{cream}\hrule width 3.5in height 0.4pt \color{black}\vspace*{5pt}} 
\setcounter{secnumdepth}{5}

\makeatletter 
\renewcommand\@biblabel[1]{#1}            
\renewcommand\@makefntext[1]% 
{\noindent\makebox[0pt][r]{\@thefnmark\,}#1}
\makeatother 
\renewcommand{\figurename}{\small{Fig.}~}
\sectionfont{\sffamily\Large}
\subsectionfont{\normalsize}
\subsubsectionfont{\bf}
\setstretch{1.125} %In particular, please do not alter this line.
\setlength{\skip\footins}{0.8cm}
\setlength{\footnotesep}{0.25cm}
\setlength{\jot}{10pt}
\titlespacing*{\section}{0pt}{4pt}{4pt}
\titlespacing*{\subsection}{0pt}{15pt}{1pt}
%%%END OF PAGE SETUP%%%

%%%FOOTER%%%
\fancyfoot{}
% \fancyfoot[LO,RE]{\vspace{-7.1pt}\includegraphics[height=9pt]{head_foot/LF}}
% \fancyfoot[CO]{\vspace{-7.1pt}\hspace{13.2cm}\includegraphics{head_foot/RF}}
% \fancyfoot[CE]{\vspace{-7.2pt}\hspace{-14.2cm}\includegraphics{head_foot/RF}}
\fancyfoot[RO]{\footnotesize{\sffamily{1--\pageref{LastPage} ~\textbar  \hspace{2pt}\thepage}}}
\fancyfoot[LE]{\footnotesize{\sffamily{\thepage~\textbar\hspace{3.45cm} 1--\pageref{LastPage}}}}
\fancyhead{}
\renewcommand{\headrulewidth}{0pt} 
\renewcommand{\headrulewidth}{0pt}
\setlength{\arrayrulewidth}{1pt}
\setlength{\columnsep}{6.5mm}
\setlength\bibsep{1pt}
%%%END OF FOOTER%%%

%%%FIGURE SETUP - please do not change any commands within this section%%%
\makeatletter 
\newlength{\figrulesep} 
\setlength{\figrulesep}{0.5\textfloatsep} 

\newcommand{\topfigrule}{\vspace*{-1pt}% 
\noindent{\color{cream}\rule[-\figrulesep]{\columnwidth}{1.5pt}} }

\newcommand{\botfigrule}{\vspace*{-2pt}% 
\noindent{\color{cream}\rule[\figrulesep]{\columnwidth}{1.5pt}} }

\newcommand{\dblfigrule}{\vspace*{-1pt}% 
\noindent{\color{cream}\rule[-\figrulesep]{\textwidth}{1.5pt}} }

\makeatother
%%%END OF FIGURE SETUP%%%

%%%TITLE, AUTHORS AND ABSTRACT%%%

\twocolumn[
  \begin{@twocolumnfalse}
\vspace{3cm}
\sffamily
\begin{tabular}{m{4.5cm} p{13.5cm} }

 &\noindent\LARGE{\textbf{Propagation of active nematic-isotropic interfaces on substrates}} \\%Article title goes here instead of the text "This is the title"
\vspace{0.3cm} & \vspace{0.3cm} \\

 & \noindent\large{Rodrigo C. V. Coelho,$^{\ast}$\textit{$^{a,b}$} Nuno A. M. Araújo,\textit{$^{a,b}$} and Margarida M. Telo da Gama\textit{$^{a,b}$}} \\\\%Author names go here instead of "Full name", etc.

&\noindent\normalsize{
Motivated by results for the propagation of active-passive interfaces of bacterial \textit{Serratia marcescens} swarms [Nat. Comm., \textbf{9}, 5373 (2018)] we use a hydrodynamic multiphase model to investigate the propagation of interfaces of active nematics on substrates.
We characterize the active nematic phase of the model and discuss its description of the statistical dynamics of the swarms. We show that the velocity correlation functions and the energy spectrum of the active turbulent phase of the model scale with the active length, for a range of activities. In addition, the energy spectrum exhibits two power-law regimes with exponents close to those reported for other models and bacterial swarms. Although the exponent of the rising branch of the spectrum (small wavevector) appears to be independent of the activity, the exponent of the decay changes with activity systematically, albeit slowly. 
We characterize also the propagation of circular and flat active-passive interfaces. We find that the closing time of the circular passive domain decays quadratically with the activity and that the structure factor of the flat interface is similar to that reported for swarms, with an activity dependent exponent. 
Finally, the effect of the substrate friction was investigated. We found an activity dependent threshold, above which the turbulent active nematic forms isolated islands that shrink until the system becomes isotropic and below which the active nematic expands, with a well defined propagating interface. The interface may be stopped by a fricion gradient. 
} 

\\%The abstrast goes here instead of the text "The abstract should be..."

\end{tabular}

 \end{@twocolumnfalse} \vspace{0.6cm}

  ]
%%%END OF TITLE, AUTHORS AND ABSTRACT%%%

%%%FONT SETUP - please do not change any commands within this section
\renewcommand*\rmdefault{bch}\normalfont\upshape
\rmfamily
\section*{}
\vspace{-1cm}

%%%FOOTNOTES%%%

\footnotetext{\textit{$^{a}$~Centro de Física Teórica e Computacional, Faculdade de Ciências,
Universidade de Lisboa, P-1749-016 Lisboa, Portugal; E-mail: rcvcoelho@fc.ul.pt}}
\footnotetext{\textit{$^{b}$~Departamento de Física, Faculdade de Ciências,
Universidade de Lisboa, P-1749-016 Lisboa, Portugal.}}

%Please use \dag to cite the ESI in the main text of the article.
%If you article does not have ESI please remove the the \dag symbol from the title and the footnotetext below.
\footnotetext{\dag~Electronic Supplementary Information (ESI) available: Description of the supplementary movies and supplementary figures. See DOI: 00.0000/00000000.}
%additional addresses can be cited as above using the lower-case letters, c, d, e... If all authors are from the same address, no letter is required

% \footnotetext{\ddag~Additional footnotes to the title and authors can be included \textit{e.g.}\ `Present address:' or `These authors contributed equally to this work' as above using the symbols: \ddag, \textsection, and \P. Please place the appropriate symbol next to the author's name and include a \texttt{\textbackslash footnotetext} entry in the the correct place in the list.}

%%%END OF FOOTNOTES%%%

%%%MAIN TEXT%%%%

\section{Introduction}

Swarming is a mode of bacterial motion on surfaces, which is very efficient in the formation of bacterial colonies or biofilms under appropriate conditions. The regulatory pathways that lead to swarming behaviour of different types of bacteria have been investigated and are reviewed in e.g. Refs.~\cite{VERSTRAETEN2008496, Beer2019}. From a biological perspective, swarming has beneficial effects in nutrient mixing and molecular transport, which promote the growth and expansion of bacterial colonies on many surfaces~\cite{121108-145434, Kohler2000}.

From a physical perspective, in swarms, individual bacteria convert energy from the environment into directed motion, which 
self-organizes into a chaotic flow known as active turbulence~\cite{Wensink14308}. Despite intensive research over decades, turbulence in isotropic fluids, described by the Navier-Stokes equation, remains an open problem in the non-linear dynamics of systems far from equilibrium.
Active turbulence is an emergent hot topic in the physics of active fluids~\cite{MarchettiRMP2013, Ramaswamy_2017}, where more complex non-linearities in the dynamics appear to play a significant role. The most obvious difference is that active turbulence occurs at very low Reynolds numbers. The statistical properties of these chaotic flows are still under debate as the characterization of active turbulence is far from trivial, both theoretically and experimentally. Recently, a numerical and theoretical analysis based on what is arguably the simplest hydrodynamic description of active fluids, revealed that the exponents that characterize the energy spectrum of active turbulence are non-universal, implying that this is a new class or set of classes of turbulent fluids~\cite{Bratanov15048}. 

A popular, if not the simplest, hydrodynamic model of active fluids is based on nematic hydrodynamics with an additional term in the equation of motion that accounts for the activity~\cite{MarchettiRMP2013, Ramaswamy_2017, Doostmohammadi2018}. This approach uses the knowledge acquired over decades of studies of the hydrodynamics of passive liquid crystals, providing theoretical tools to incorporate the effect of the boundaries and of relevant particle parameters, such as flow alignment or shape, on the collective flow~\cite{10.1039/C8SM02103A, C9SM00859D, Doostmohammadi2016}. 
Multiphase or multicomponent nematic hydrodynamic models have also been used to characterize the structure and dynamics of interfaces between the active nematic and passive phases. Recently, we used a multiphase model to investigate the interfaces of confined active liquid crystals. The study was restricted to narrow channels and low activities where the nematic phase is ordered. At very low activities we found stable interfaces that propagate at constant velocity both for extensile and contractile systems, while at higher activities an interfacial instability was observed for extensile nematics~\cite{C9SM00859D}.

Recently, experiments with swarms of \textit{Serratia marcescens} have shown that exposure to ultraviolet light can temporarily or permanently passivate the bacteria, depending on the time and intensity of the light~\cite{yang2019quenching, Patteson2018}. By using different screens, almost circular or flat active-passive interfaces have been created. The experiments revealed that these interfaces propagate, with rich dynamics that have been characterized quantitatively. 

%-----------------figure-------------------------------
\begin{figure*}[th]
\center
\includegraphics[width=\linewidth]{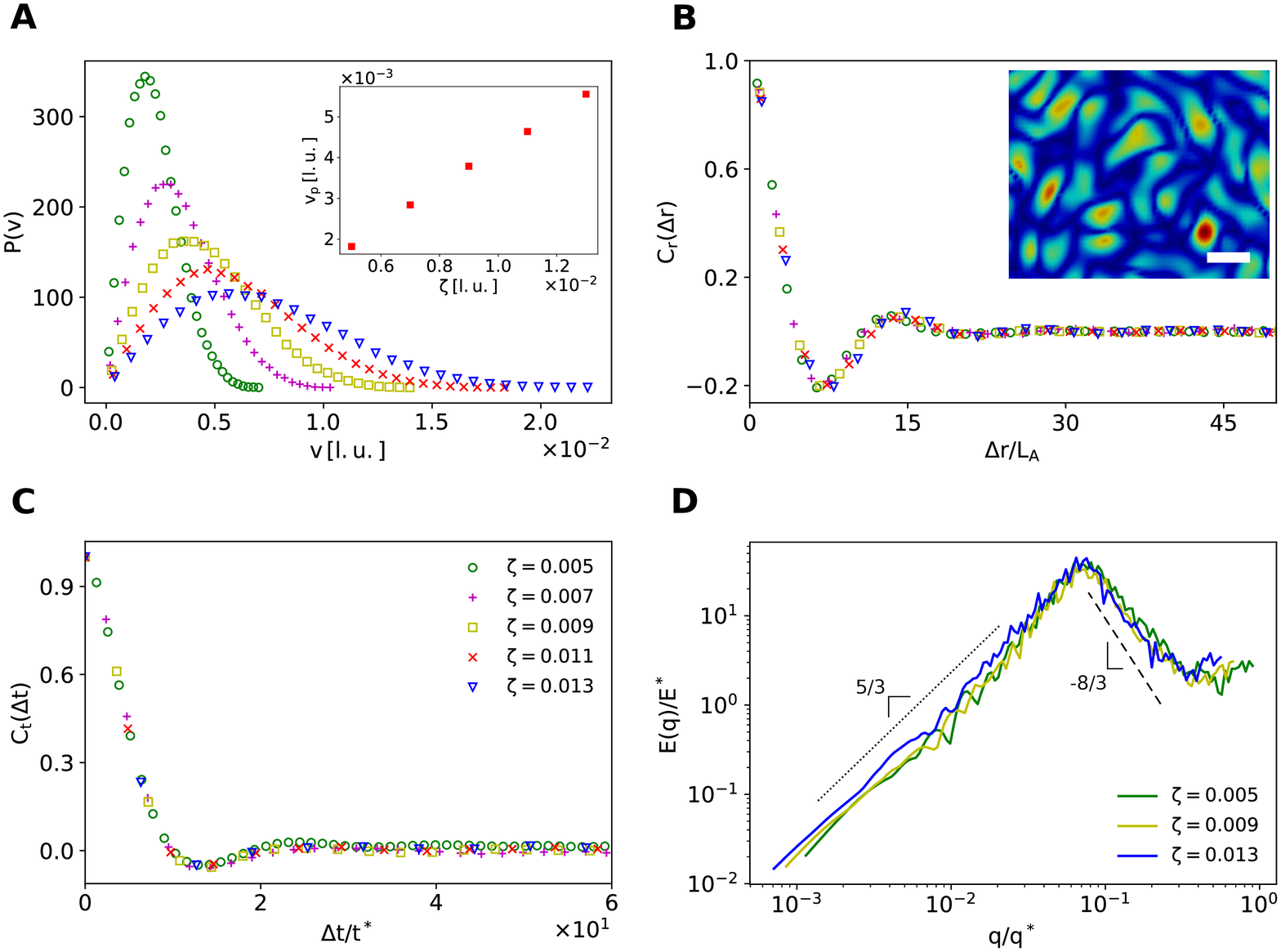}
\caption{Statistical properties of the active turbulent nematic phase for a substrate with friction $\chi=0.1$ and different activities. A) Normalized velocity distribution for five activities (legend as in C). The inset shows the peak velocity $v_p$ for each distribution, which is used as the characteristic velocity in the next panels. B) Space correlation function calculated using the vorticity field for five activities (legend as in C); The inset illustrates the vorticity field in a small window for $\zeta = 0.007$, where the scale bar is $10$ l.u. (see Fig. 1 of ESI for the vorticity and velocity field in the entire system). C) Time autocorrelation function calculated using the vorticity field for five activities. D) Energy spectrum for three different activities. Notice that the correlation functions and the energy spectrum collapse when reduced units (non-dimensional variables) are used.}
\label{bulk-fig}
\end{figure*}
%------------------------------------------------------

Motivated by these experimental results on the propagation of active-passive interfaces of \textit{Serratia marcescens}, we study the propagation of nematic-isotropic interfaces of turbulent active nematics on surfaces. The model is based on active nematic hydrodynamics where the activity is coupled to the local nematic order parameter, with an additional term that accounts for the
substrate friction. As in our previous work, we solve the hydrodynamic equations of active nematics numerically using a hybrid method of lattice-Boltzmann and finite differences ~\cite{C9SM00859D, 2019arXiv190713415C}. 

We set the model parameters so that the system is in the active turbulent phase and check that the dynamical statistical 
properties of this phase are consistent with the results reported for the swarms. In addition, we investigate the effect of the 
activity and find that, for the range of activities under consideration, both the correlation functions and the energy spectrum 
scale with the active length and the characteristic velocity in the active nematic phase. We then characterize the interfacial dynamics and confirm that the model describes semi-quantitatively the dynamics reported for the propagation of the interfaces of \textit{Serratia marcescens} swarms. In particular, the structure factor of the flat interface follows approximately a power-law as reported in the experiments, with an exponent that depends on the activity. Finally, we investigate the effect of substrate friction 
and find that there is a friction threshold above which the active nematic phase retracts, suggesting a mechanism to stop the interfacial propagation by means of a friction gradient. 

The paper is organized as follows. In Sec.~\ref{method-sec}, the equations of motion of a multiphase active nematic are briefly described. In Sec.~\ref{nematic-sec} the dynamical statistical properties of the turbulent active nematic phase are calculated, namely the space and time vorticity correlation functions and the energy spectrum. In Sec.~\ref{nematic-sec}, the propagation of circular and flat interfaces is investigated for a range of activities. In Sec.~\ref{flat-sec}, the effect of substrate friction is studied. In Sec.~\ref{multicomponent-sec}, we discuss the applicability of a multicomponent model to describe the propagation of the interfaces discussed previously and illustrate the differences from the results of the multiphase model used in the previous sections. Details of the multicomponent model are found in the Appendix. In Sec.~\ref{conclusions-sec}, we make the concluding remarks.

\section{Method}
\label{method-sec}

%-----------------figure-------------------------------
\begin{figure*}[thb]
\center
\includegraphics[width=\linewidth]{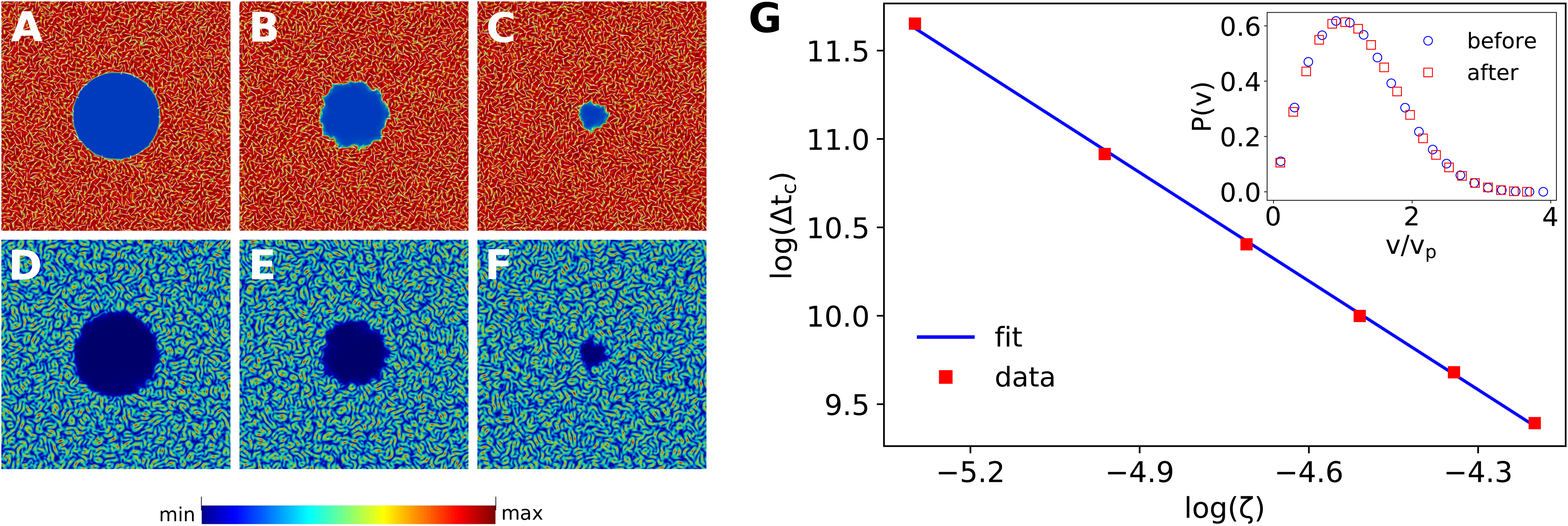}
\caption{Closing time for a circular interface. A to F show the fields for a system with activity $\zeta = 0.009$ and substrate friction $\chi=0.1$ at three different instants of reduced time: (A and D) $t_1/t^* = 186.9$, (B and E) $t_2/t^* = 215.7$, (C and F) $t_3/t^* = 269.6$ . A, B and C) Order parameter, where the colors represent the scalar order parameter (red for nematic and blue for isotropic) and the white lines indicate the directors. D, E and F) Velocity, where the colors represent the magnitude of the local velocity (from $5.09\times 10^{-7}$ to $1.44\times 10^{-2}$). G) Closing time for different activities. The linear fit gives the slope $-2.00 \pm 0.02$. The inset shows the velocity distribution in the active turbulent phase of a system with $\zeta =0.009$ before passivating the circular domain  ($t_b/t^* = 176.1$) and after its closure ($t_a/t^* = 1078.4$).}
\label{closing-fig}
\end{figure*}
%------------------------------------------------------

We model the dynamics of the swarms using a continuum hydrodynamic theory for active nematic liquid crystals. This is based on the equilibrium Landau-de Gennes theory for non-uniform nematics also known as the $Q$ tensor theory~\cite{p1995physics}. 
The uniaxial tensor order parameter is, $Q_{\alpha \beta} = S(n_\alpha n_\beta - \delta_{\alpha \beta}/3)$, where $S$ is the scalar order parameter and $n_\alpha$ is the director. $Q$ is traceless and symmetric and describes the nematic ordering tensor that 
may vary in space and time, while $S$ measures the degree of ordering and $n$ its direction. Symmetry between $n$ and $-n$ ensures nematic ordering. For simplicity, we assume that $Q$ is uniaxial, which is exact in bulk nematics and an excellent approximation under most conditions. 

The temporal evolution of the uniaxial tensor order parameter $Q$ is given by the Beris-Edwards equation~\cite{p1995physics}:
\begin{align}
  \partial _t Q_{\alpha \beta} + u _\gamma \partial _\gamma Q_{\alpha \beta} - S_{\alpha \beta}(W_{\alpha\beta}, Q_{\alpha\beta}) = \Gamma H_{\alpha\beta} , 
  \label{beris-edwards}
\end{align}
where $\Gamma$ is the system dependent rotational diffusivity and $u_\alpha$ is the velocity field. The co-rotational term is given in terms of the vorticity and the strain rate tensors by: 
\begin{align}
 &S_{\alpha \beta} = ( \xi D_{\alpha \gamma} + W_{\alpha \gamma})\left(Q_{\beta\gamma} + \frac{\delta_{\beta\gamma}}{3} \right) 
 + \left( Q_{\alpha\gamma}+\frac{\delta_{\alpha\gamma}}{3} \right)(\xi D_{\gamma\beta}-W_{\gamma\beta}) \nonumber \\& 
 \quad\quad\:\:\:\:- 2\xi\left( Q_{\alpha\beta}+\frac{\delta_{\alpha\beta}}{3}  \right)(Q_{\gamma\epsilon} \partial _\gamma u_\epsilon), \\
 &W_{\alpha\beta} = (\partial _\beta u_\alpha - \partial _\alpha u_\beta )/2, \quad \text{and} \quad D_{\alpha\beta} = (\partial _\beta u_\alpha + \partial _\alpha u_\beta )/2.\nonumber
\end{align}
The flow depends on a particle dependent parameter $\xi$, which is positive for rod-like and negative for disk-like particles, while the molecular field $H_{\alpha\beta}$ describes the relaxation of the order parameter towards equilibrium:  
\begin{align}
 H_{\alpha\beta} = -\frac{\delta \mathcal{F}}{\delta Q_{\alpha\beta}} + \frac{\delta_{\alpha\beta}}{3} \Tr \left( \frac{\delta \mathcal{F}}{\delta Q_{\gamma \epsilon}} \right).
\end{align}
The Landau-de-Gennes free energy for a nematic in three-dimensions is given by:
\begin{align}
\mathcal{F} =& \int_V \,d^3 r \left[ \frac{A_0}{2}\left( 1- \frac{\gamma}{3} \right) Q_{\alpha \beta}^2 - \frac{A_0\gamma}{3} Q_{\alpha \beta} Q_{\beta \gamma} Q_{\gamma \alpha}\right.  \nonumber \\ &+ \left. \frac{A_0\gamma}{4} (Q_{\alpha \beta} Q_{\alpha \beta} )^2 + \frac{K}{2} (\partial _\gamma Q_{\alpha \beta})^2.\right],
\end{align}
where $A_0$ is a system dependent positive constant, $K$ is the elastic constant in the single-constant approximation and $\gamma$ is a parameter that determines the phase transition (e.g., temperature). The first three terms correspond to the bulk free energy and the last is the elastic energy, i.e., the energy cost associated to distortions with respect to the uniform alignment of the director field. The isotropic ($S=0$) and the nematic ($S=S_N$) phases coexist at $\gamma = 2.7$ where the scalar order parameter is $S_N=1/3$. The nematic correlation length is given by, $L_N = \sqrt{27 K / (A_0 \gamma)}$, and sets the scale for the decay of nematic order. As the nematic-isotropic transition is first-order, the nematic correlation length is finite at the transition and of the order of the characteristic length of the particles. This length sets the scale of the nematic-isotropic interfacial width and the size 
of the defect cores in the nematic phase~\cite{p1995physics, chaikin2000principles}.
In the simulations that follow, we fix $\gamma$ at the passive nematic-isotropic coexistence. The elastic constant and the bulk nematic parameters are also fixed, and thus $L_N$ is fixed.

The velocity field is governed by the continuity and Navier-Stokes equations:
\begin{align}
&\partial_t \rho + \partial_\beta (\rho u_\beta) = 0 \nonumber \\
&\partial_t (\rho u_\alpha) + \partial _\beta  (\rho u_\alpha u_\beta) = -\chi u_\alpha + \partial_\beta [\eta(\partial_\alpha u_\beta + \partial_\beta \partial_\alpha) \label{navier-stokes-eq}\\
&\quad\quad\quad\quad \quad\quad\quad \quad\quad\quad\: + \Pi_{\alpha\beta} -\zeta Q_{\alpha\beta} ].  \nonumber 
\end{align}
where $\rho$ is the density. The first term on the right hand side of the Navier-Stokes equation describes the friction with the substrate with strength set by $\chi$. The second term describes viscous stresses with $\eta$ the shear-viscosity. The last term is the active stress with $\zeta$ the activity parameter, which is positive for extensile systems and negative for contractile ones. 
Finally, the passive nematic stress tensor is:
\begin{align} 
 \Pi_{\alpha\beta} =& -P_0 \delta_{\alpha\beta} + 2\xi \left( Q_{\alpha\beta} +\frac{\delta_{\alpha\beta}}{3} \right)Q_{\gamma\epsilon}H_{\gamma\epsilon} \nonumber \\ &- \xi H_{\alpha\gamma} \left( Q_{\gamma\beta}+\frac{\delta_{\gamma\beta}}{3} \right) - \xi \left( Q_{\alpha\gamma} +\frac{\delta_{\alpha\gamma}}{3} \right) H_{\gamma \beta} \nonumber \\ &- \partial _\alpha Q_{\gamma\nu} \,\frac{\delta \mathcal{F}}{\delta (\partial_\beta Q_{\gamma\nu})} + Q_{\alpha\gamma}H_{\gamma\beta} - H_{\alpha\gamma}Q_{\gamma\beta} ,
 \label{passive-pressure-eq}
\end{align}
where $P_0=\rho/3$ is the hydrostatic pressure (for the D3Q19 lattice) and the other terms derive from the free-energy functional.
The system dependent parameter $\xi$ sets the tendency of the flow to align the particles in the flow direction. 

We used the numerical method described in Refs.~\cite{C9SM00859D, 2019arXiv190713415C}. This method uses lattice-Boltzmann to solve the equations for the fluid motion, recovering Eq.~\ref{navier-stokes-eq} macroscopically, and finite differences to solve Eq.~\ref{beris-edwards}. Some of our results are given in lattice units expressed as ``l.u.'' in the figures: the distance between nodes is $\Delta x = 1$ and the time step is $\Delta t = 1$ (see for instance Refs.~\cite{ThampiEPL2015, Doostmohammadi2018} to transform these to physical units). In the following simulations (except in Sec.~\ref{multicomponent-sec}), we set the parameters: $\tau=1.2$ (relaxation time in the Boltzmann equation, which gives the kinematic viscosity $\nu=[\tau-1/2]/3 = 0.23$), $K=0.01$, $A_0=0.1$, $\gamma=2.7$, $\Gamma=0.34$ and $\rho=40$ on average. The aligning parameter is set to $\xi=0.7$ (flow aligning regime). With these parameters $L_N= 1$ l.u. meaning that we do not resolve smaller length scales. We note that the length of \textit{Serratia maracescens} is $5-7\,\mu$m~\cite{Patteson2018}, seting $L_N$ in the experiments to a few $10 \mu$m. The transition from flow aligning to flow tumbling regimes occurs at $\xi^*= 3 S/(S+2) = 0.43$ if $S=1/3$ (coexistence $\gamma$). The remaining parameters are given in the text or in the figures of the corresponding simulations.

%-----------------figure-------------------------------
\begin{figure*}[thb]
\center
\includegraphics[width=0.95\linewidth]{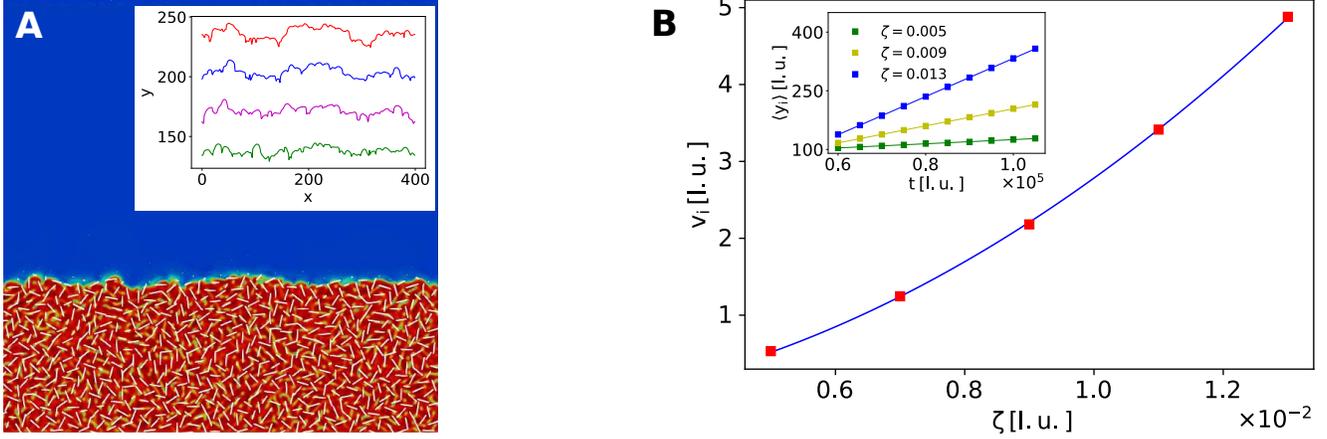}
\caption{Propagation of a flat interface. A) Typical order parameter field ($\zeta = 0.009$ and $t/t^* = 251.6$) for an interface propagating upwards. The blue and red correspond to the isotropic and the active nematic phases while the white lines indicate the directors. The inset shows the interface at four different instants of reduced time: 251.6, 305.5, 359.5, 413.4. B) Interface velocity $v_i$ for systems with different activities. The squares represent the measurements from the simulations and the solid line is a quadratic fit: $f(x) = ax^2 +b$, where $a=30.2 \pm 0.2$ and $b= (-24 \pm 2)\times 10^{-5}$. As illustrated in the inset, the interface velocity is determined by a linear fit to the average interface position $\langle  y_i \rangle$ as a function of time.}
\label{flat-fig}
\end{figure*}
%------------------------------------------------------

\section{Characterization of the turbulent active nematic phase}
\label{nematic-sec}

In this section, we analyze the statistical properties of the turbulent active nematic phase of the model and compare them with those reported for the swarms of \textit{Serratia marcesens}~\cite{Patteson2018}. We consider a two-dimensional simulation box with dimensions  $L_X \times L_Y = 400 \times 400$, and set the friction between the active nematic and the substrate $\chi=0.1$. The active liquid crystal is initialized at rest with $S=1/3$ and directors pointing in random directions at each point. We choose the activity in a range where the system relaxes to the active turbulent state after a few thousand iterations. We take measurements after $10^5$ time steps.

In addition to the nematic correlation length $L_N$ and the length of the simulation box $L_X$, which will be fixed in what follows, there are two other length scales that are relevant and will be varied. The first is the active length, $L_A = \sqrt{K/\zeta}$, the ratio of  elastic and active stresses, which drives spontaneous elastic distortions and hydrodynamic flow. The second is the friction or screening length, $L_F = \sqrt{\nu/\chi}$, the ratio of the shear stress and that due to substrate friction, which sets the scale over which frictional damping removes kinetic energy from the system~\cite{PhysRevX.5.031003, stone_ajdari_1998}. In order to observe fully developed active turbulence, these lengths must satisfy $L_N < L_A, L_F \ll L_X$. The second inequality is required to avoid finite-size effects while the first ensures that the creation of defects is not hindered by the fact that the defect cores, set by $L_N$, are larger than the active length, $L_A$~\cite{C6SM00812G} (see Fig. 2 of ESI). In this section the screening length is set to $L_F=1.52$, while the active length is varied in the range $0.88 <  L_A  < 1.41$. This choice of parameters, at the border of fully developed active turbulence, was taken by visual inspection of the swarm dynamical behavior reported in Ref.~\cite{Patteson2018}. A quantitative comparison of the statistical dynamical behavior of the model and that \textit{Serratia marcescens} swarms is described in the next paragraphs.

We start by calculating the distribution of velocities for systems with different activities, as shown in Fig.~\ref{bulk-fig}A. The curves resemble Maxwell-Boltzmann distributions in 2D, in line with the results reported for the bacterial swarms~\cite{Patteson2018}. Note that as the activity increases the peak velocity also increases. The dependence is approximately linear, for the range of activities considered, as shown in the inset of Fig.~\ref{bulk-fig}A, where the peak velocity $v_p$ is ploted as a function of the activity. In the active turbulent regime the characteristic velocity is expected to scale with $v_{ch} \sim \sqrt{\alpha}$~\cite{C6SM00812G}. The reason why we observed peak velocities that scale linearly is most likely due to the narrow range of activities considered.

We note that this increase in the peak velocity contrasts with the effect of the activity in promoting local nematic order above the passive nematic-isotropic transition reported earlier and highlights the fact that the non-equilibrium active turbulent phase cannot be described by an equilibrium system at an effective temperature~\cite{ThampiEPL2015}.

Next, we calculate the normalized space correlation function of the vorticity field, $\boldsymbol{\omega}=\nabla \times \mathbf{u}$:
\begin{align}
C_r(\Delta r) = \left \langle  \frac{\boldsymbol{\omega}(r_0)\cdot \boldsymbol{\omega}(r_0+\Delta r)}{\vert \boldsymbol{\omega}(r_0)  \vert^2} \right \rangle,
\end{align}
where the average is over the reference position $r_0$ and time. The vorticity is calculated using fourth-order finite-differences: $\frac{\partial f}{\partial x} (x) = [-f(x+2\Delta x) +8 f(x+ \Delta x) - 8 f(x-\Delta x) + f(x-2\Delta x)]/(12\Delta x) + \mathcal{O}(\Delta x^4)$. In Fig.~\ref{bulk-fig}B, we plot the correlation function versus the position reduced by the active length, $L_A$, for systems with different activities. The active length is the ratio of the elastic and active stresses and sets the scale of the vortices in the active nematic phase. We note that the curves collapse into a single curve revealing that the active length is dominant in this regime. This result implies that even for the system with the largest activity, where the active length is smaller than the nematic correlation length, the generation of defects is not significantly hindered. The space correlation functions of the velocity field (see Fig. 3 of ESI) exhibit a similar behaviour with a slightly different characteristic length. The scaling with $L_A$ is also similar.   

We define the characteristic length of the system as the position of the first zero of the correlation function, $C_r(\Delta r) = 0$,  which is found to be $L_{ch}\approx 4.5 L_A$. The average size of the vortices $L_v$ is estimated from the position of
the second zero of the correlation function, which is: $L_v \approx 11.4 L_A$. 
We note that, for this range of activities, the space correlation function of the model is similar to that reported for the swarms of \textit{Serratia marcescens}, which also exhibits a well defined minimum. As the activity of the swarms is likely to change linearly through exposure to small doses of ultra-violet light~\cite{yang2019quenching}, the position of the minimum could be used to estimate the active length.

The normalised time autocorrelation function of the vorticity field,
\begin{align}
C_t(\Delta t) = \left \langle  \frac{\boldsymbol{\omega}(t_0)\cdot \boldsymbol{\omega}(t_0+\Delta t)}{\vert \boldsymbol{\omega}(t_0)  \vert^2} \right \rangle,
\end{align}
for different activities also collapses, see Fig.~\ref{bulk-fig}C, when the time is reduced by the active time defined as $t^* = L_A/v_p$. This is the minimal time required for a particle to move a distance equal to an active length. We define the characteristic time of the system as the time of the first zero of the correlation function, $C_t(\Delta t) =0$, which is found to be $t_{ch}\approx 10.3  L_A /v_p$. This minimum is less well defined both in the simulations and in the experiments~\cite{Patteson2018} and it is therefore less likely to be useful to estimate the activity of the swarms. 

For completeness, we calculate the energy spectrum of the active nematic hydrodynamic model with substrate friction,
\begin{align}
E(q) = \frac{q}{2\pi} \int d \mathbf{R}\, e^{-i \mathbf{q}\cdot \mathbf{R}} \langle \mathbf{v}(t,\mathbf{ r_0})\cdot  \mathbf{v}(t, \mathbf{r_0}+\mathbf{R})  \rangle,
\end{align}
where the average is over time and space. The energy spectrum is plotted in Fig.~\ref{bulk-fig}D for three different activities. The reduced wavevector is $q/q^*$ with $q^*=2\pi/L_A$ and 
the reduced energy is $E/E^*$ with $E^* = v_p^2/2$. 

A prominent feature of the spectra is the presence of a peak at $q_p \approx 0.44 L_A^{-1}$, close to $ 2 \pi/ L_v$, where $L_v$ is the average size of the vortices given by the position of the second zero of the space correlation function. As for the correlation functions, the energy spectra nearly collapse when reduced units are used.
Within the limited range of wavectors that were simulated, the energy increases with $q/q^*$ for two decades, up to a well defined peak and then decreases for a decade before leveling off. This leveling off at large wavevectors, corresponds to length scales of the order of 4 l.u. and it is reasonable to assume that it is due to the ultraviolet cut-off. We note that this cut-off sets in at scales that
are a couple of times larger than the microscopic length scales of the system. The infrared cut-off ocurs at scales of the order 
of the lateral size of simulation box. 

The first thing to notice is that the energy spectrum resembles that observed experimentally~\cite{Patteson2018}. Both the model and the experimental spectra, however, exhibit large fluctuations and although the measured exponents are similar, a quantitative analysis reveals some differences. For systems with activities $0.005$, $0.009$ and $0.013$, we measured exponents $1.79 \pm 0.03$, $1.70 \pm 0.03 $ and $1.70 \pm 0.02$ reasonably in line with the exponent $5/3$ reported for the increasing branch of the spectrum, while we measured $-1.96\pm 0.07$, $-1.98 \pm 0.08$ and  $-2.3 \pm 0.1$ that contrast with the exponent $-8/3$ reported for the decreasing branch of the energy spectrum of \textit{Serratia marcescens} swarms~\cite{Patteson2018}.

At the same level of analysis the results for the energy spectra obtained from the Q tensor model are similar to those reported in Ref.~\cite{Wensink14308} for Bacillus subtilis suspensions in a quasi-2D geometry, as well as to the results obtained from a minimal continuum model for incompressible bacterial flow, which we call the v-model (for velocity). Beyond the inertial 
terms, the Q tensor and v- models have different non-linearities and this agreement appears somewhat fortuitous.   
In fact, a subsequent study of the v-model, based on sophisticated numerical techniques, revealed~\cite{Bratanov15048} that the exponents of the rising branch of the energy spectrum are non-universal. Furthermore, the exponents obtained for the range of parameters corresponding to the active turbulent regime rule out the value of $5/3$ for the v-model reported earlier.~\cite{Wensink14308}. Although this does not necessarily rule out the exponents reported here for the Q tensor model, it does show that computing reliable energy spectra in the active turbulent regime requires the use of specific numerical techniques, which in view of what is known from numerical studies of turbulence in isotropic fluids is not surprising.

Although active turbulence remains a very open problem, the Q tensor theory describes qualitatively the statistical dynamical 
properties of bacterial swarms. In what follows, we use this model to investigate the propagation of active-passive interfaces. 

% Energy spectrum
% Growing
% zeta = 0.005: a = 1.79 +/- 0.03
% zeta = 0.009: a = 1.70 +/- 0.03
% zeta = 0.013: a = 1.70 +/- 0.02
% 
% Decreasing
% zeta = 0.005: a = -1.96 +/- 0.07
% zeta = 0.009: a = -1.98 +/- 0.08
% zeta = 0.013: a = -2.3 +/- 0.1

\section{Propagation of the interface}
\label{propagation-sec}

In this section, we analyze the propagation of an isotropic-nematic interface on a substrate (with friction $\chi = 0.1$) for a small range of activities. We start by investigating the dynamics of the closing of a circular interface and proceed with the propagation of flat interfaces. Both of these problems are relevant in the study and control of biofilms~\cite{yang2019quenching, Patteson2018}. 

\subsection{Circular interface}
\label{circular-sec}

In order to simulate a circular interface, we initialize the system as described in Sec.~\ref{nematic-sec}: a domain with dimensions $L_X \times L_Y = 400 \times 400$ at the passive nematic-isotropic transition with  $S=1/3$ and directors pointing in random directions. We employ periodic boundary conditions in both directions. After $50000$ iterations, when the system is in the steady-state active turbulent regime, we create a circular interface by imposing $S=0$ in a circle of radius $R=75$ for $2000$ iterations to mimic the irradiation used in the experiments to passivate a closed domain~\cite{Patteson2018, yang2019quenching}. As the active force in the model is proportional to the local gradient of the nematic order parameter (see Eq.~\ref{navier-stokes-eq}), the fluid becomes almost static in this circular region, see Fig.~\ref{closing-fig}A and D. After removing the constraint, the circular interface starts moving inwards and the isotropic region shrinks with time until it closes. At this point the active turbulent phase - which is the steady state for this set of parameters - fills the simulation box. In the inset of Fig.~\ref{closing-fig}G, we show that the velocity distribution before the interface is formed and after its closure is unchanged, as in this model the order parameter is not conserved. A quantity of interest is the closure time, $\Delta t_c$, which is plotted in Fig.~\ref{closing-fig}G in a logarithmic scale for different activities. The linear relation suggests a power-law dependence with exponent $\alpha=-2$. This is probably the simplest way to measure the bacterial activity, provided that the swarm dynamics is described by the model (as suggested in the previous section) and a reference activity is defined.

\subsection{Flat interface}
\label{flat-sec}

%-----------------figure-------------------------------
\begin{figure}[t]
\center
\includegraphics[width=\linewidth]{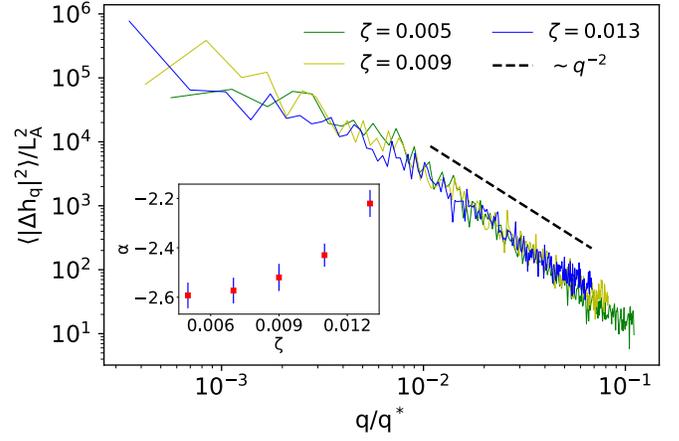}
\caption{Interfacial structure factor: temporal average of the magnitude squared of the Fourier modes of the interfacial height fluctuations, $ \Delta h (x,t) = y_i(x,t)-\bar y_i(t)$, where $\bar h(t)$ is the average interfacial height at time $t$. The inset shows the measured exponent $\alpha$ of the power-law decay, for each activity. The error bars are the errors of the linear fits.}
\label{structure-fig}
\end{figure}
%------------------------------------------------------

For the flat interface, the system is initialized as in Sec.~\ref{circular-sec}, with different boundary conditions. The boundaries are now periodic in the horizontal direction and no-slip at the top and bottom. The scalar order parameter is fixed at the values of the isotropic $S=0$ (top) and nematic $S=1/3$ (bottom) coexisting phases, with zero gradient for the directors in the bottom (zero order extrapolation from the first fluid neighbor). After the active turbulent 
steady state is reached, an isotropic region is created by imposing the condition $S=0$ for a period of time from $t=50000$ to $t=52000$ in the region $y>L_Y/4$. 

The resulting interface is flat on average and propagates into the isotropic region. At time $t$, we measure the interfacial height $y_i$ at each position $x$, through the scalar order parameter field $S(x,y,t)$. Starting from the top, we identify the interfacial height with the value of $y$ where $S=S_N/2$. In Fig.~\ref{flat-fig}A, we plot the order parameter field and in the inset the interfacial height $y_i (x)$ at four different times. 

We found that the interface moves steadily with the same (statistical) shape: flat on average with an interfacial width, $w$, that remains constant with time, in line with the experimental observations~\cite{Patteson2018} (see Fig. 4 of ESI). If the system is wide enough ($L_F,\,L_A \ll L_X$), $w$ does not depend on $L_X$.  The interfacial width at time $t$ is 
the standard deviation of the interfacial height fluctuations about the mean, $\Delta h (x,t) = y_i(x,t)- \bar y_i (t)$, where 
$\bar y_i(t) = \langle y_i (x,t) \rangle$ and the angular brackets denote an average over the horizontal coordinates. The variance $w^2$ may be written as:
\begin{align}
w^2 (t) = {1 \over L_X} \int _0 ^{L_X}  \Delta h(x,t) ^2 dx = \sum_q \vert \Delta h_q(t) \vert^2
\end{align}
where $q=n\pi /L_X$ and the Fourier mode with wavevector $q$ is, 
\begin{align}
\Delta h_q(t) = \frac{1}{L_X} \int _0 ^{L_X} \Delta h (x, t) e^{-iqx} dx.
\end{align}

The inset of Fig.~\ref{flat-fig}B, shows that the mean interfacial height, $ \bar y_i(t)$, increases linearly with time implying that the interfacial velocity is constant. This linear propagation  is in line with the experimental observations at early times. At longer times, however,  the experimental results suggest that the interface propagates diffusively, which is to be expected if the order parameter is conserved. In Sec.~\ref{multicomponent-sec}, we use a multicomponent model to describe the interfacial propagation of a system where the order parameter is conserved. Although this model captures the diffusive interfacial motion, other features of the swarm interfacial dynamics are not described by the model.     

The interface velocity $v_i$, in Fig.~\ref{flat-fig}B, was found to increase quadratically with the activity, in line with the quadratic behavior of the closing time reported in Fig.~\ref{closing-fig}. Indeed, if the velocity of the circular interface is independent of the radius, we expect $\Delta t_c \sim \zeta^{-\alpha}$, with $\alpha$ the exponent of the flat interfacial velocity, $v_i \sim \zeta^\alpha$.
 
In thermal equilibrium, the creation of an interface costs an excess free energy proportional to the interfacial tension, $\sigma$, and the length of the interface, $L$. In standard capillary wave theory, the increase in interfacial length arising from the long-wavelength interfacial height fluctuations is~\cite{doi19840880621}
\begin{align}
\Delta L  =  \frac{L_X}{2}   \sum_q q^2 \vert \Delta h_q \vert^2.
\end{align}
Furthermore, the equilibrium interfacial height mode amplitudes are given by the equipartition theorem, and 
\begin{align}
\vert \Delta h_q \vert^2 = {2 kT \over L_X \sigma q^2}
\end{align}
implying that the interfacial structure factor $\vert \Delta h_q \vert^2$ decays at small wavectors as $1/q^2$. 

Here, we define the interfacial structure factor, by the time average of the magnitude squared of the Fourier modes of the interfacial height fluctuations. This is ploted in Fig.~\ref{structure-fig} for three different activities. As for the other quantities analysed previously, we find that the curves nearly collapse, when the units are reduced by the active length. The $1/q^2$ decay expected at equilibrium was reported in the swarm experiments~\cite{Patteson2018}. Over the limited range of wavevectors that were simulated, we observe power-law decays with exponents that are close but are not $-2$ (they vary from $-2.6$ for $\zeta = 0.005$  to $-2.2$ for $\zeta = 0.013$). As shown in the inset of Fig.~\ref{structure-fig}, the exponents obtained from the model actually depend on the activity. We note that both the theoretical and the experimental structure factors are rather noisy and the power-law behaviour, let alone the exponent, is suggested rather than firmly established.    

% zeta 0.005: 63.9/LA^2
% zeta 0.009: 46.8/LA^2
% zeta 0.013: 31.8/LA^2

% Structure factor:
% zeta = 0.005, a = -2.59 +/- 0.05
% zeta = 0.007, a = -2.57 +/- 0.05
% zeta = 0.009, a = -2.41 +/- 0.05
% zeta = 0.011, a = -2.43 +/- 0.05
% zeta = 0.013, a = -2.22 +/- 0.06

\section{Effect of friction}
\label{friction-sec}

%-----------------figure-------------------------------
\begin{figure}[t]
\center
\includegraphics[width=\linewidth]{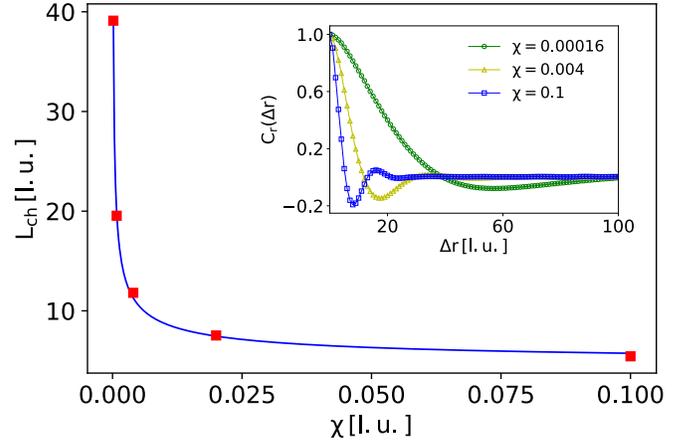}
\caption{Characteristic length $L_{ch}$ as a function of substrate friction, defined as the first zero of the space correlation function,  $C_r(\Delta r)=0$ (inset). The squares are the data and the solid line is a fit of the form $L_{ch} = a \chi^{-\frac{1}{2}}+b$, with $a=0.44$, and $b=4.35$.}
\label{friction-fig}
\end{figure}
%------------------------------------------------------
%-----------------figure-------------------------------
\begin{figure*}[thb]
\center
\includegraphics[width=\linewidth]{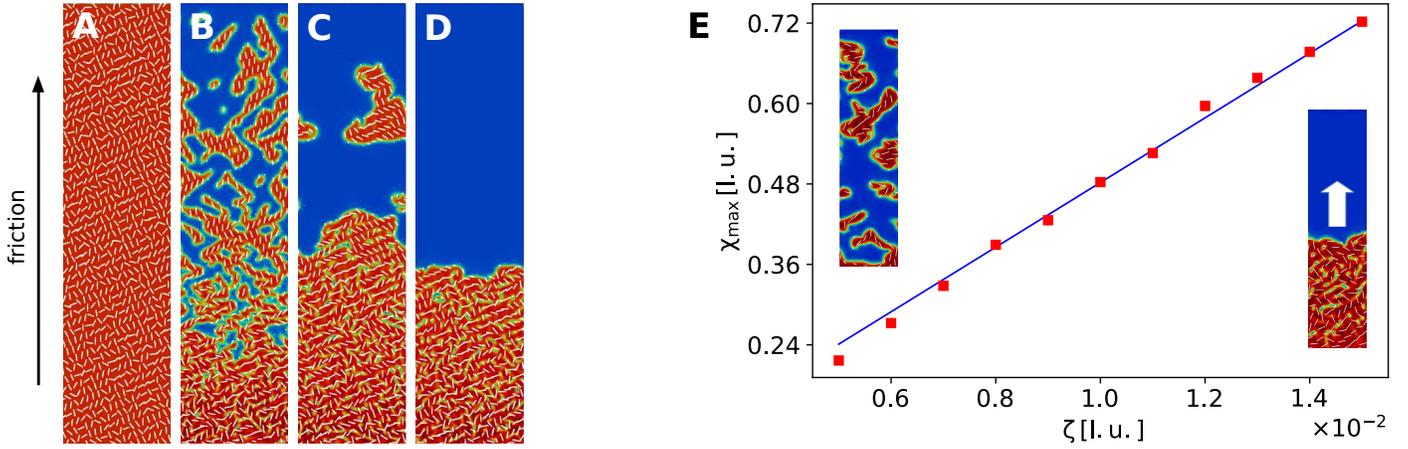}
\caption{Active nematic on a substrate with a gradient of friction at different instants of time. The friction varies linearly from $f=0.1$ at the bottom to $f=0.9$ at the top. The red and blue colors represent the nematic and the isotropic phases while the white lines represent the directors. A) $t/t^* = 0$, B) $t/t^* = 18.0$, C) $t/t^* = 359.5$, D) $t/t^* = 3594.6$. E) Friction threshold as a function of the activity. The squares are the data and the solid line is a linear fit $f(x)=a\,x$, with $a=48.2 \pm 0.4$. Above the line the active turbulent nematic retracts and the isotropic phase is the steady state while below the nematic expands and the active turbulent phase is the steady state. The insets depict typical snapshots of the transients above ($\zeta=0.009$, $\chi=0.5$) and below the coexistence line ($\zeta=0.009$, $\chi=0.4$).}
\label{grad-frict-fig}
\end{figure*}
%------------------------------------------------------

We now turn to the effect of the substrate friction. We fix the activity at $\zeta=0.009$ and investigate the effect of changing the friction.

We start by investigating how the characteristic length $L_{ch}$ depends on the substrate friction. The results are shown in Fig.~\ref{friction-fig}, where $L_{ch}$ is defined as the first zero of the space correlation function of the vorticity field (inset of Fig.~\ref{friction-fig}). $L_{ch}$ is expected to depend linearly on the screening length: $L_{F}=\sqrt{\nu/\chi}$, where $\nu$ is the kinematic viscosity~\cite{Doostmohammadi2018}. The results of a fit $L_{ch} = a \chi^{-\frac{1}{2}}+b$ to the data points confirm 
the expectation. The non-zero value of $b$ depends on the particular definition of $L_{ch}$.  The screening length was varied from $L_F=1.5$ at $\chi=0.1$ to $L_F=37.9$ at $\chi=0.00016$. In the upper limit of the range we observed finite-size effects although $L_F$ is only about one tenth of $L_X$. However, the size of the vortices is greater than $100$ l.u. as shown in the inset of Fig.~\ref{friction-fig}. The energy spectra for systems with different substrate friction follow approximately the $5/3$ and $-8/3$ power-laws with the peak changing with the friction (see Fig. 5 of ESI). 

We found that when the friction is above a threshold that depends on the activity, the turbulent active nematic forms clusters that shrink with time until the domain is filled with the isotropic phase. This is illustrated in the insets of Fig.~\ref{grad-frict-fig}E: on the left ($\chi=0.5$ and $\zeta=0.009$), the active nematic clusters disappear at long times and the steady state is isotropic; on the right ($\chi=0.4$ and $\zeta=0.009$), the active nematic domain expands as the interface propagates upwards and the steady state is active turbulent nematic. For a system with activity $\chi=0.009$ the friction threshold was found to lie between $0.4$ and $0.5$. 

In order to calculate the friction threshold for a range of activities, we used a gradient of friction, which describes a substrate with properties (e.g., rugosity or agar concentration) that change along a certain direction. We use the same boundary conditions as in Sec.~\ref{flat-sec}, $L_X \times L_Y = 100 \times 400$, and the substrate friction is taken to vary linearly from $\chi=0.1$ at the bottom to $\chi=0.9$ at the top. We initialize the system with active nematic and directors pointing in random directions. A typical time evolution is shown in Fig.~\ref{grad-frict-fig}A to D. The active nematic forms clusters in the region with higher friction and remains 
as a single domain in the region where the friction is lower. At long times, the clusters shrink and eventually disappear and the interface becomes almost flat and pinned, at a fixed position, in the steady-state regime. We determine the friction threshold $\chi_{max}$ as the substrate friction at the position where the interface is pinned. 

We checked that this is indeed the friction threshold by simulating substrates with uniform friction around this value. We found
that the friction of the uniform substrate, which separates the shrinking and expanding active nematic regimes agrees with the friction where the interface is pinned to within $5\%$. The friction threshold as a function of the activity is plotted in Fig.~\ref{grad-frict-fig}E. This line may be interpreted as the coexistence phase diagram of isotropic and active turbulent phases on uniform substrates with friction.    

We failed to observe static interfaces with a global change of the fields. The static or pinned interface was observed only in the presence of gradients. In addition to the gradient of friction at fixed activity and $\gamma$, it is possible to observe static interfaces in a gradient of activity, at fixed friction and $\gamma$ or even in a gradient of $\gamma$, at fixed activity and friction, suggesting that these transitions are continuous rather than first-order. 

\section{Conserved order parameter}
\label{multicomponent-sec}

%-----------------figure-------------------------------
\begin{figure*}[thb]
\center
\includegraphics[width=\linewidth]{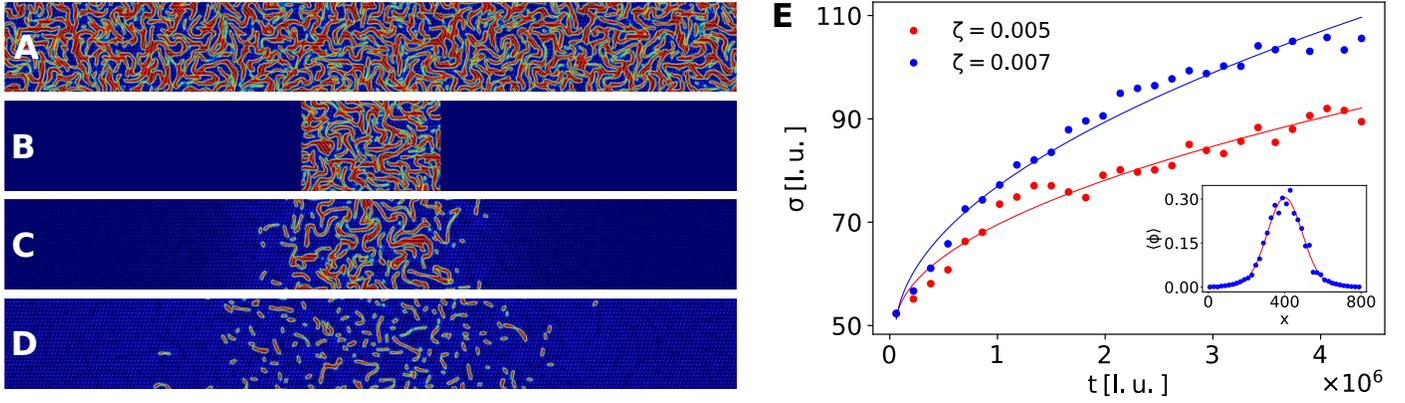}
\caption{Evolution of the order parameter in the multicomponent model for $\zeta=0.007$. A) $t=4\times 10^4$. B) $t=5\times 10^4$, when the interface is created. C) $t=5\times 10^5$. D) $t=5\times 10^6$.  D) Temporal evolution of the standard deviation in the spatial distribution of the order parameter for two different activities. The circles represent the measurements from the simulations and the solid lines represent a curve fit of the form $\sigma(t) = a\sqrt{t} + b$. The inset illustrates the measurement of the standard deviation for $z=0.007$ and $t = 1.5 \times 10^6$, where the points are the data from the simulation and the solid line is the fitted Gaussian. }
\label{multicomponent-fig}
\end{figure*}
%------------------------------------------------------

To address the diffusive interfacial motion suggested by the experiments on swarms, we considered the interfacial dynamics of an active nematic multicomponent hydrodynamic model (see the Appendix and Ref.~\cite{PhysRevLett.113.248303} for details). The multicomponent model describes an active emulsion of two immiscible fluids: the active nematic and a passive isotropic fluid. There are two main differences between the multicomponent model considered previously and this one. The most important difference is that the order parameter is conserved, that is the dynamics conserves the amount of nematic and isotropic fluids.  
Secondly, in the previous model, we restricted the system to a 2D geometry but the model is 3D and thus the directors could stick
out of the plane, while the multicomponent model is truly 2D. This is not a relevant difference as in the simulations reported in the previous section the directors are confined almost exclusively within the plane.   

In the multicomponent model the local concentration of nematic, varies from $\phi=0$ to $\phi=1$, which are the concentrations of the isotropic and nematic phases, respectively. We initialize the system, $L_X \times L_Y = 800 \times 100 $, with homogeneous concentration $\phi=0.5$ and directors pointing randomly. We allow the system to evolve and, in the interval $50000 < t < 51000$, we set $\phi = 0$ and $S=0$ in the region $x<325$ and $x>475$. We employ periodic boundary condition in both directions. 

One obvious difference between the results of the two models is that the interface of the multicomponent model is not well defined as illustrated in Fig.~\ref{multicomponent-fig}A to D. At $t=50000$, the separation between the isotropic fluid and the active nematic is clear, but, at later times, filaments of active nematic disperse into the isotropic fluid and it becomes difficult to identify an interface. In Fig.~\ref{multicomponent-fig}E, we show the evolution of the standard deviation of the order parameter, which was measured by fitting a Gaussian to the spatial distribution of the concentration $\phi$ (see the inset). The curve is a fit of the form $\sigma(t) = a\sqrt{t} + b$, showing that the nematic diffuses into the isotropic phase and that this diffusion depends on the activity. In the experiments with swarms of  \textit{Serratia marcescens}~\cite{Patteson2018}, the authors suggest that the position of the interface, at late times, follows a $\sqrt{t}$, suggesting the diffusion of activity as described by the multicomponent model. The 
experimental results, however, are not very conclusive although they seem to indicate the existence of two regimes. It is not 
difficult, however, to fit a $\sqrt{t}$ to the early time and a linear fit to the late time regime. In the early time regime the interface 
is actually broader and then it becomes sharper with a shape that appears to remain constant as it propagates. Clearly, further research is needed to clarify these results both theoretically and experimentally. 

\section{Summary and conclusion}
\label{conclusions-sec}

We have used numerical simulations to investigate the propagation of nematic-isotropic interfaces of active nematics subject to friction between the fluid and the substrate. We used the active nematic hydrodynamics multiphase model, based on the Q tensor theory 
in 3D, confined in 2D on a substrate with friction.  

We have shown that the correlation functions and the energy spectrum of the active turbulent nematic phase, in a small range of activities, nearly collapse if the units are reduced by the active length. The energy spectrum exhibits a peak at a wavector that corresponds to the size of the vortices. Both the correlation functions and the energy spectrum are found to be in semi-quantitative agreement with the results reported for swarms of \textit{Serratia marcescens}~\cite{Patteson2018}. 

The closing time of a circular interface was calculated for different activities and it was found to decay quadratically with the activity, which may provide an easy way to measure the activity of the swarms. For the flat propagating interface, the interface velocity was found to be constant as the order parameter is not conserved in the multiphase model. This model should provide an even better description if the bacteria are temporarily passivated becoming active during the time of the experiment. An alternative multicomponent model that conserves the order parameter did not improve the overall agreement with the experiments on swarms. In particular it failed to describe the propagation of well defined interfaces, which were observed experimentally~\cite{Patteson2018}. We did not consider the reproduction of bacteria in the multicomponent model, but, when this is relevant, a growth term may be included in the model as reported in~\cite{C9SM01210A}. We note that a growth term is an alternative, albeit different, way to describe models with non-conserved order-parameters.

We measured the interface velocity for different activities and calculated the structure factor of the propagating interface, which was found to deviate from the results expected for interfaces at equilibrium. We note that for interfaces of active Brownian particles that undergo motility induced phase separation, the interfacial structure factors were reported to follow, roughly, the equilibrium (equipartition) $1/q^2$ decay~\cite{PhysRevLett.95.268301, C8SM00899J}. In our model the interfaces propagate, which may be a significant difference from the static interfaces of active brownian particles. In addition, the results for the interfacial structure factor are rather noisy and much larger simulations would be required to exclude the equilibrium behaviour of the interfacial structure factor of the Q tensor model.    

We also investigated the effect of substrate friction on the interfacial propagation. We verified that the characteristic length of the active turbulent nematic scales linearly with the screening length as expected. Interestingly, we found a threshold above which the active nematic phase forms clusters that shrink until the system becomes isotropic. Below this friction threshold, the interface propagates and the active nematic phase expands until it fills the domain. By using a gradient of friction, the interface 
can be pinned and become static. We also found that the gradient of any field (friction, activity or temperature) may be used to pin the interface, which may be of some relevance in the control of biofilm formation.

\section*{Conflicts of interest}
There are no conflicts to declare.

\section*{Acknowledgements}

We acknowledge financial support from the Portuguese Foundation for Science and Technology (FCT) under the contracts: PTDC/FIS-MAC/28146/2017 (LISBOA-01-0145-FEDER-028146) and UID/FIS/00618/2019.

\section*{Appendix: Multicomponent model}

To simulate the interface of a model with a conserved order parameter, we implemented the multicomponent model of Ref.~\cite{PhysRevLett.113.248303}, summarized below.

The system consists of a mixture of a nematic and an isotropic fluid and the total mass of each fluid is conserved. The relative concentration is given by the scalar parameter $\phi$, which is $1$ for the nematic and $0$ for the isotropic fluid. The order parameter of the nematic phase is, $Q_{\alpha\beta} = S(2 n_\alpha n_\beta - \delta_{\alpha\beta})$, where $n_\alpha$ is the director and $S$ is the scalar order parameter. The free nergy density of the system is:
\begin{align}
f =& \frac{1}{2} A\phi^2 (1-\phi)^2 + \frac{1}{2}C \left( S^2_n\phi - \frac{1}{2}Q_{\alpha \beta} Q_{\alpha \beta}  \right)^2 \nonumber \\ 
&+ \frac{1}{2} K_1 \partial _\gamma \phi \partial _\gamma \phi + \frac{1}{2}K_2 \partial _\gamma Q_{\alpha \beta} \partial _\gamma Q_{\alpha \beta},
\end{align}
where $A$, $C$, $K_1$ and $K_2$ are positive constants. The first two terms, which describe the bulk free energy, have two minima: the nematic phase with $S=S_N$ and $\phi = 1$ and the isotropic phase with $S=0$ and $\phi=0$. The chemical potential $\mu$ is a Lagrange multiplier that conserves the total amount of each fluid:
\begin{align}
\mu = \frac{\partial f}{\partial \phi} - \partial _ \gamma \left( \frac{\partial f}{\partial (\partial_\gamma \phi)}  \right).
\end{align}
The order parameter and the concentration evolve according to the Beris-Edwards and the Cahn-Hilliard equations:
\begin{align}
&\partial_t \phi + \partial_\beta (\phi u_\beta) = M \nabla^2 \mu\\
&(\partial_t+u_\kappa  \partial_\kappa ) Q_{\alpha\beta} = -\xi \Sigma _{\alpha\beta\kappa\lambda}D_{\kappa\lambda} - T_{\alpha\beta\kappa\lambda}W_{\kappa\lambda} + \Gamma H_{\alpha\beta},
\end{align}
where $M$ is the mobility constant which controls the diffusion and:
\begin{align}
& \Sigma _{\alpha\beta\kappa\lambda} = S^{-1}_n Q_{\alpha\beta} Q_{\kappa\lambda} - \delta_{\alpha\kappa} (Q_{\lambda\beta}+S_N\delta_{\lambda\beta}) \nonumber \\
&\quad \quad \quad \:\:\:\:  -(Q_{\alpha\lambda}+S_N\delta_{\alpha\lambda})\delta_{\kappa\beta} + \delta_{\alpha\beta}(Q_{\kappa\lambda}+S_N\delta_{\kappa\lambda}), \\
&T_{\alpha\beta\kappa\lambda} = Q_{\alpha\kappa} \delta_{\beta \lambda} - \delta_{\alpha\kappa} Q_{\beta\lambda}.\nonumber
\end{align}
The molecular field gives the relaxation of the order parameter towards the minimum of the free energy:
\begin{align}
H_{\alpha\beta} = \frac{1}{2}(\delta_{\alpha\beta}\delta_{\kappa\lambda} - \delta_{\alpha\kappa}\delta_{\beta\lambda}-\delta_{\alpha\lambda}\delta_{\beta\kappa})\left[ \frac{\partial f}{\partial Q_{\kappa\lambda}} - \partial_\lambda \left( \frac{\partial f}{\partial (\partial_\gamma  Q_{\kappa\lambda})}  \right)  \right]
\end{align}
The density and velocity field are governed by the continuity and Navier-Stokes, Eq.~\ref{navier-stokes-eq}, equation with the stress tensor given by:
\begin{align}
&\Pi_{\alpha\beta} = -P_0\delta_{\alpha\beta}+(\xi \Sigma_{\alpha\beta\kappa\lambda} + T_{\alpha\beta\kappa\lambda})H_{\kappa\lambda}\nonumber \\
&+(f-\mu\phi) \delta_{\alpha\beta} - \frac{\partial f}{\partial (\partial_\beta \phi)} \partial_\alpha\phi - \frac{\partial f}{\partial (\partial_\beta Q_{\kappa\lambda})}\partial_\alpha Q_{\kappa\lambda} .
\end{align}

In the simulations, we have used the following parameters: $S_N=1$, $A=0.08$, $C=0.5$, $K_2=0.005$, $K_1=0.01$, $\Gamma=0.1$, $M=0.1$, $\tau=1.2$ (relaxation time used in the Boltzmann equation), $\chi=0.1$, $\xi = 0.7$, $\rho=40$ on average.

%%%END OF MAIN TEXT%%%


\balance


%%%REFERENCES%%%
\bibliography{rsc} %You need to replace "rsc" on this line with the name of your .bib file
\bibliographystyle{rsc} %the RSC's .bst file

\end{document}